\documentclass[times, 10pt,twocolumn]{article} 
\usepackage{latex8}
\usepackage{times}
\usepackage{epsf}
\usepackage{amsfonts}

\newcommand{\cH}{{\cal H}}
\newcommand{\cA}{{\cal A}}

\def\C{{\mathbb{C}}}

\def\R{{\mathbb{R}}}

\begin{document}

\title{Are there quantum bounds on the recyclability \\
of  clock signals in low power computers?}

\author{Dominik Janzing and Thomas Beth\\
Institut f\"{u}r Algorithmen und Kognitive Systeme\\ Universit\"{a}t Karlsruhe  \\ 
Am Fasanengarten 5\\ D-76 131 Karlsruhe\\ Germany\\
}

\maketitle
\thispagestyle{empty}

\begin{abstract}
Even if a logical  network consists of thermodynamically reversible
gate operations, the computation process may have  high dissipation
 rate
if the gate implementation is controlled by external 
clock signals. It is an open question whether the global 
clocking mechanism necessarily envolves irreversible processes.
However, one can show that it is not possible 
to extract any timing information from a  micro-physical clock
without disturbing it. 
Applying recent results of quantum information theory
we can show a hardware-independent lower bound on the timing information
that is necessarily destroyed if one tries  to copy the signal. 
The bound becomes tighter for low energy signals, i.e.,
the timing information gets more and more quantum.
\end{abstract}

%------------------------------------------------------------------------- 
\Section{Introduction}

To invent new methods of low power 
computation is an important
goal of research. In the middle future,
the physical limits of miniaturization and 
reduction of power consumption will play
a crucial role. 
The fact that thermodynamic laws give  constraints   
to the energy consumption in information processing, has already 
been recognized by Landauer in the sixties \cite{Landau}. 
The statement, that whenever the information $n$ bits are lost during the 
computation process the energy $\ln 2 \, n kT$ is dissipated 
(where $k$ is Boltzmann's constant and $T$ is the absolute temperature)
is nowadays known as {\it Landauer's principle}.
The fact that any logical function 
can be embedded in a logically reversible network by
Toffoli gates \cite{Toffoli} shows that
Landauer's principle does not give any obvious 
lower bounds for the dissipation during the computation process.
However, this argument is  not sufficient to show 
that computation without
dissipation is possible. Even if the logical gates act 
thermodynamically reversibly 
on the register the implementation may consume energy
if the process is controlled by an external clock signal.
However, the quantum cellular automaton described in \cite{Marg}
indicates that this problem can in principle be avoided by
doing computation without refering  to a global clocking  mechanism.

Here
we consider the problems that may appear if one wants to keep the
usual concept of global clocking with a clock signal
that should not be 
 `used up'  when it controls 
the gate implementations. The key problem is to what extent
timing information can be read out without destroying it, i.e.,
to what extent timing information is 
{\it quantum} information that is impossible to clone.

We assume the clock signal to be given by
the state of a  micro-physical system represented by a quantum density matrix.
In Section \ref{disturb} we show that at every moment where the clock
signal controls another physical system it will necessarily
be disturbed by the system's back-action. Using results of modern
quantum information research we show that it is impossible in principle
 to extract information about the actual time from a micro physical clock 
(represented by a finite dimensional density matrix)
without disturbing the clock's state.
Generalizations to infinite dimensional systems and quantitative
statements on this result are subject of further research.

However, we can give quantitative statements in a slightly different 
situation. If one tries not only to extract {\it some} timing information
from the clock but wants to copy as much timing information as possible to 
another system one necessarily destroys some timing information
in the original clock.
This partial destruction of the original clock signal gets more and more 
relevant if low energy signals are considered. These bounds are
sketched in Section \ref{Fisher}. The proofs can be found in \cite{clock}.

\section{Every clock is disturbed when it is read out}

\label{disturb}

Let $\rho_t$ be the finite dimensional 
density matrix of the clock at time $t$.
Since it is assumed to be a closed physical system it is evolving according
to the evolution
\[
\rho_t = \exp (-iHt)\, \rho \,\exp(iHt)
\]
where $H$ is the clock's Hamiltonian.

For the following reason it is not possible to extract any information about $t$ without disturbing
the state $\rho_t$ if no prior knowledge about $t$ is given.
Consider two states $\rho_{t_1}$ and $\rho_{t_2}$. Following
 \cite{KI} we conclude that a measurement distinguishing
between those states 
can be implemented without changing the states 
if and only if the following condition holds:
Let $\oplus \cH_j =\cH$ be the decomposition of the clock's Hilbert space
into those maximal subspaces that are invariant under the action of
 $\rho_{t_1}$ and $\rho_{t_2}$.
Let $A_j$ and $B_j$ be the corresponding block matrices of $\rho_{t_1}$ and
$\rho_{t_2}$, respectively.
Then there exists a number $j$ such that $tr(A_j)\neq tr(B_j)$.
An appropriate observable for the distinction is for example 
the projection
onto the Hilbert space $\cH_j$.

This gives a rule for constructing disturbance free 
measurements if we are sure 
that either $t=t_1$ or $t=t_2$ by prior knowledge.
If nothing is known about the time we have to find common invariant subspaces
  of all $\rho_t$. Those have clearly to be invariant under the action of $H$.
But the trace of the block matrices 
on those dynamically invariant subspaces is conserved. There is hence
no way to gain information about $t$ without disturbing the system.
The derivation of quantitative bounds on minimal disturbance is a 
difficult task. The information about $t$ has to be quantified
as well as the disturbance.

Note that the statement that there is no extraction of information
about $t$ relies essentially on the continuity of $t$.
The necessary and sufficient condition above allows to construct systems where
a discrete set of states $\rho_{t_1}, \rho_{t_2},\dots$ can indeed be 
distinguished without disturbing the system.
This shows that if an external clock tells us that the actual 
time $t$ is in the set $\{t_1,t_2,\dots \}$ we can read out our "`discrete
clock' without disturbing it.
Remarkably, the fact that readout without disturbance is in principle 
possible in the discrete case
shows the advantage of controlling information processing by
external clock signals: Let now $\rho_t$ be the density matrix 
of any micro physical device. Use the clock signal to switch on an interaction
between the considered device and another system.
Then the sequence of states $\rho_{1},\rho_{2},\dots$ at the times
$t_1,t_2,\dots$   can control the 
other system without back-action (see Fig.1).

\begin{figure}
\centerline{
\epsfbox[0 0 132 125]{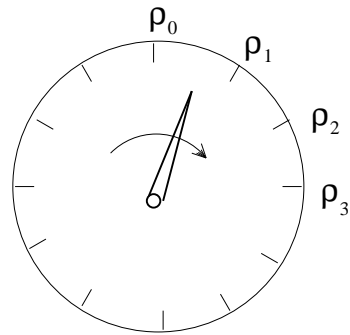}}
\caption{Only at specific times can a quantum clock be read out without 
being disturbed. Hence the read out process has to be controlled by a 
`meta clock'.}
\end{figure}

The relevance of the above observations for the reusability of clock signals
has to be subject of further discussions. Clearly the gate has 
extracted some 
information about the actual time from the clock signal when it is triggered.
This results in a disturbance of the clock signal's quantum state.
However, one may find systems where this extracted timing information 
flows back to the signal {\it after} the implementation is performed.
But as long as the triggered implementation is still running
the gate has some timing information and the clock signal's state 
is still disturbed.

\section{Copying a clock signal or its timing information}

\label{Fisher} 

In the last section we have shown that we cannot extract timing information
without disturbing the clock signal. However, we were not able to
make quantitative statements about the tradeoff between information 
gain and disturbance. In the following situation, we can find
quantitative results:
It is natural to ask whether it is possible to read out all the information
about the actual time that is contained in the signal, i.e., to copy it. 
In the following we will specify precisely what it means
to `copy the timing information'.
We use {\it Fisher timing information}, a quantity that is actually 
well-known in a more general context of estimating parameterized
quantum  or classical statistical states
\cite{BC94,BCM96,Hole,Hel,Cra}. In \cite{clock}  it is used
 in the specific context 
of investigating the quality  of clocks, explicitly using the terminology
`Fisher timing information'.
One might think of the Fisher timing information $F$ of a system
with statistical states $\rho_t$ as the quotient $1/(\Delta t)^2$
if $\Delta t$ is the error for the optimal estimation of $t$ 
that is achievable
by measuring the state. The correct and formal definition
of $F$ is given in the appendix.
But the rough explanation of $F$ can be taken literally
in many simple examples: 
assume $\rho_t$ to be pure quantum states with a Gaussian energy distribution
with standard deviation $\Delta E$. Then $F= 4 (\Delta E)^2$
(where we have measured the energy in the unit $\hbar$)
and    
one can easily construct measurements allowing an estimation of $t$
with standard deviation $1/\sqrt{F}$ and no better estimation can exist
due to Heisenberg's uncertainty relation.
Another example where $F$ has a simple intuitive meaning 
is the following.
Consider a classical pulse, i.e., a classical quantity $f$ that changes
in time according to the function $t\mapsto f(t)$.
This system has infinite timing information for any non-trivial function
$f$ since the determination of $f(t)$ allows to distinguish between
$t$ and $t'$ for arbitrarily small $t-t'$.
If we introduce an unknown time delay of the signal with
Gaussian statistics the timing information of the `smeared out' signal 
is the standard deviation $\Delta t$ of the delay (see Fig.2).
One might think of $f(t)$ as the classical current or voltage of any device
or even the intensity of a  (classical) light field.
One can equivalently think of a classical signal
moving with the velocity $v$. If the exact position of
the signal is unknown according to a Gauss distribution with
uncertainty $\Delta x$ the timing information is given by 
$v^2/(\Delta x)^2$.

\begin{figure}
\centerline{
\epsfbox[0 0 126 91]{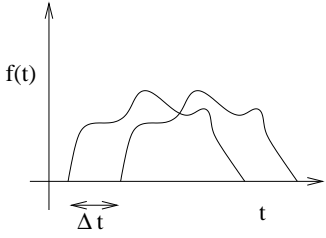}
}
\caption{Classical signal having a well-defined shape but unknown time 
delay with
standard deviation $\Delta t$.}
\end{figure}

If $\rho_t$ is a finite dimensional density matrix of a quantum system 
evolving according to the Hamiltonian $H$, the quantity $F$ is
less simple to calculate and is given by
\[
F=tr (\dot{\rho} \Gamma^{-1} \dot{\rho})\,,
\] 
where  $\dot{\rho}:=i[H,\rho]$ and $\Gamma^{-1}$ is the pseudo-inverse
of the super-operator $\Gamma$ with $\Gamma a:= 1/2 (\rho a +a \rho)$
acting on the set of self-adjoint matrices \cite{BC94,BCM96,Hole,Hel,clock}.

In \cite{clock} we derived the following quantum bound for
copying timing information:

Assume a signal with Fisher timing information $F$ enters a device 
(an amplifier for example) and triggers two outgoing signals with
Fisher information $F_1$ and $F_2$, respectively.
Then one has 
\begin{equation} \label{ent}
1/F_1 + 1/F_2 \geq 2/F + 2/\langle E^2\rangle\,,
\end{equation}
where $E$ is the total energy of the {\it outgoing} signals
and 
$\langle E^2\rangle$ is the expectation value of the square
of this energy.
Accordingly, the
 time uncertainties $\Delta t_1$
and $\Delta t_2$   of the 
2 outgoing signals triggered by an in-going signal
with time uncertainty $\Delta t$  satisfy the following inequality: 
\[
(\Delta t_1)^2 + (\Delta t_2)^2 \geq 2 (\Delta t)^2 + 2/\langle E^2\rangle \,.
\]
Trying to give both signals the same timing information, i.e.,
$\Delta t_1=\Delta t_2$, one obtains
\[
(\Delta t_1)^2 \geq (\Delta t)^2 + 1/\langle E^2\rangle \,.
\]
We conclude that low energy signals loose part of their timing information 
when
they are copied  (see Fig.3).

\begin{figure}
\centerline{
\epsfbox[0 0 232 135]{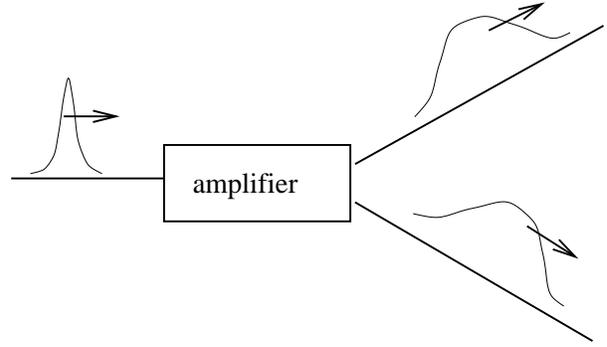}
}
\caption{The outgoing signals are less localized in time then the input.
The functions can be interpreted as the probability distribution of the 
signal's time of arrival at a certain point. The clock consisting
of  both outputs
cannot be better than the clock defined by the input (with respect 
to the quasi-order sketched in Section \ref{quasi}).}
\end{figure}

In \cite{clock} we have constructed an example
showing that $\Delta t_1= \Delta t_2 
=\Delta t$ is indeed possible in the limit $\Delta E \to \infty$.
Note that this does {\it not} imply  that a high amount of energy is 
{\it dissipated} when the signal is copied. The energy has only to 
be {\it available}. 
Whether the disturbance of the clock signal (as explained in Section
\ref{disturb}) results necessarily
in energy consumption, is unclear but if the signal looses some of its
timing information it is unclear how to run a reversible process if the
clock signal is included in the consideration.

\begin{figure}
\centerline{
\epsfbox[0 0 177 113]{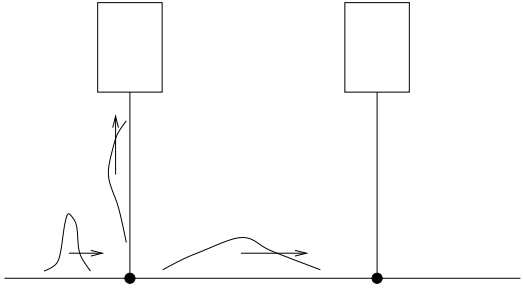}
}
\caption{When the clock signal controls a device some of the signal's
timing information is {\it copied}. The possibility of copying timing 
information is restricted by fundamental quantum bounds independent
from the hardware.}
\end{figure}

Note that whenever the clock signal controls a device part of the
signal's timing information is copied. This is illustrated in Fig.4.

\section{New thermodynamical constraints by a quasi-order of clocks}

\label{quasi}

The proof of inequality (\ref{ent}) can be found in \cite{clock}
and relies on a formal concept
called {\it quasi-order of clocks}
classifying physical systems
(quantum or classical) with respect to their 
timing information. Here timing information is not to be understood
in the sense of a single quantity 
but refers to the statistical distinguishability of the 
quantum or classical states $\rho_t$ at different times $t$.
So to speak, it classifies the quality of clocks.
This quality has many aspects: 
The physical  system $A$  can be better than the system  $\tilde{A}$ 
with respect to the distinguishability the states 
at the times $t_1$ and $t_2$
and nevertheless $\tilde{A}$ may be better than $A$ with respect to the
distinguishability between the states at the times 
 $t_3$ and $t_4$.
We write $A\geq \tilde{A}$ if $A$ is not worse than $\tilde{A}$ 
with respect to any criterion.
The physical meaning is that is is possible in principle to realize 
a process with input $A$ and output $\tilde{A}$ such that the process
has not to be controlled by an external clock.
This idea can be formalized by describing 
the process by completely positive maps \cite{Kraus}. The 
fact that no external clock is allowed to run the process 
corresponds to  the condition that this map has to be covariant
with respect to the time evolution of $A$ and $\tilde{A}$.
For instance, if $H$ and $\tilde{H}$ are the Hamiltonians of the input and 
output quantum system, a process  $G$ 
is a completely positive trace-preserving  map 
from the set of possible density matrices of the system $A$ to the set of
density matrices of $\tilde{A}$. It can be implemented without referring
to external clocks if it satisfies the covariance  condition
\[
G([H,.])=[\tilde{H}, G(.)]\,.
\]
The condition states that the process is the same 
if it is implemented at a later time since it does not matter
whether we let the {\it input system} $A$ evolve for the time $t$ 
and apply then the process $G$
or we apply $G$ first and then wait for the time $t$ such that the 
{\it output system} $\tilde{A}$ 
is evolving in time.

The output clock is not better than the input clock. As shown in \cite{clock}
this principle gives constraints to many physical processes.
Remarkably, the covariance condition also appeared in
a rather different context in \cite{JWZ00}.
There we found thermodynamic constraints on processes that can be run
with a negligible amount of implementation energy.

\Section{Relative timing information and synchronization}

The considerations above refer to clocks showing the {\it absolute}
time. At first sight one might think that the relevance of these results
is restricted, since computation relies rather on the fact that gate 
implementations have to be synchronized and not that the gates are implemented
at specific times. Hence one will rather be interested in {\it relative timing
information} than in {\it absolute timing information} discussed in the 
previous sections. However, as we have shown in \cite{clock},
the problem of measuring the {\it quality of synchronization} of two
signals can be reduced to the problem of measuring the localization
in time of a single signal. We have introduced a formal concept
called {\it quasi-order of synchronism} that is shown to be 
mathematically isomorphic to the quasi-order of clocks.

\Section{Why the signal energy is relevant}

The fact that our bound for copying timing information
depends on the {\it signal energy}, might be astonishing.
The tradeoff between information gain about a quantum state
and the disturbance of the system does not refer a priori 
to the energy of the quantum system \cite{Fu96b}.
Nevertheless it is easy to get a rough idea why
energy is a relevant quantity if the {\it timing} information of a system
should be copied:
An unknown quantum state $\rho \in \rho_1,\rho_2,\dots$ 
can be perfectly copied (`broadcasted') if and only if
all the density matrices $\rho_i$ commute (see \cite{Fu96} for the proof
and the definition of {\it broadcasting} quantum states).
Consider now the states $\rho_t$ of the Hamiltonian time evolution. 
It is easy to see that all the states $\rho_t$ commute if and only if
the time  evolution is trivial, i.e. all states are the same.
But it may be possible to find times $t_1,t_2,\dots$ such that
all the states $\rho_{t_i}$ commute and  we can copy the states 
perfectly if the prior information $t \in \{t_1,t_2,\dots \}$ is given.
For systems with high energy spread these times can be arbitrarily close
together. An example is a pure quantum state given
by the equal superposition 
\[
|\psi\rangle :=\frac{1}{\sqrt{n}}\sum_{j\leq n} |j\rangle
\]
where $j$ is an energy eigenstate with energy $jE$. 
Then the states at the times  $t= j2 \pi E/ n$ 
are mutually orthogonal for different $j$, i.e., they
can be copied without disturbance. The average  energy  of the system 
is $E n/2$, i.e., the distinguishable states get closer and closer
together (compare \cite{ML98}) for $n\to\infty$.
Even if no prior information about $t$ is given, one can imagine that
it is possible to extract some information about $t$ while 
disturbing the state only a little bit as long as one wants to
know $t$ only up to an error that is much larger than $E/ n$.
One can sketch the idea of these remarks by claiming that
the timing information in systems with low energy is essentially
quantum information and timing information in systems with high energy spread
can have a high part of classical information that can be extracted
without disturbing the system too much.

The result has an interesting implication for 
low power computation: consider
a fanout  in low power circuits. Then the two out-going signals
cannot have the same localization in time as the in-going signal. 

These results suggest the following problem of extremely 
low power computation: if one uses low power clock signals
it is difficult to copy the signal and distribute it
to many devices. If the signals  contain more energy the question of
the reusability  becomes more relevant.

\section{Conclusions}

We have shown that every clock signal, as far as it is given by a finite 
dimensional quantum system is necessarily 
disturbed when it controls a device.
Quantitative results concerning the tradeoff  
between the effect of the clock on the network 
and the back-action  are the subject of further research.
A first step towards such a quantitative analysis
shows that strong disturbance of the signal 
is inevitable if most of the signal's timing information
is transfered to the triggered system.
Our bounds on the disturbance become relevant if the
signal energy is in the order of $\hbar /\Delta t$ where
$\Delta t$ is the signal's accuracy in time.
To what extent this result implies bounds on
energy dissipation for all computation processes with a global clocking
mechanism
is unclear. It is not even clear how to define
such a mechanism formally. 
However,  it indicates serious difficulties that may appear if 
the usual concept of clocking 
is maintained in future low power technology.

\section*{Acknowledgements}

\noindent  
Thanks to Michael Frank for an interesting email discussion.
His  statements encouraged us to figure out the relevance of former results
for the reusability of clock signals.

This work has been supported by DFG-grants of the `Schwerpunktprogramm
Verlustarme Informationsverarbeitung'.

\section*{Appendix}

The
Fisher timing information $F$ of a quantum or classical system is
defined as follows \cite{clock}. Let $\cA$ be the (unital) $C^*$-algebra of observables
of the system \cite{BR1} and $\rho$ be the system's state, i.e., a
positive funtional from $\cA$ onto the set $\C$ with $\rho(1)=1$.
Let $(\alpha_t)_{t\in \R}$ be the time evolution of the system, i.e.,
a strongly continuous one parameter group of automorphisms on $\cA$.
Then we define $F$ as
\[
F:= \sup_{A}\frac{(d/dt \, \rho (\alpha_t(A)))^2}{\rho (A^2)}
\]
where the supremum is taken over all self-adjoint $A\in \cA$.

\end{document}